\documentclass[pra,aps,11pt]{revtex4}
\usepackage{graphicx}

\begin{document}
\title{Influence of Phase Matching on the Cooper Minimum in Ar High Harmonic Spectra}

\author{J. P. Farrell, L. S. Spector, B. K. McFarland, P. H. Bucksbaum, M. G\"uhr}
\address{
Stanford PULSE Institute, Stanford University, Stanford, CA 94305, USA\\
Chemical Science Divison, SLAC National Accelerator Laboratory, Menlo Park, CA 94025, USA\\
Departments of Physics and Applied Physics, Stanford University, Stanford, CA 94305, USA\\
\email[]{E-mail: mguehr@stanford.edu}
}
\author{M. B. Gaarde, K. J. Schafer}
\address{
Department of Physics and Astronomy, Louisiana State University, Baton Rouge, LA 70803-4001 USA\\
}

\begin{abstract}
We systematically study the influence of phase matching on interference minima in high harmonic spectra. We concentrate on structures in atoms due to interference of different angular momentum channels during recombination. For this purpose, we use the Cooper minimum (CM) in argon at $\sim 47$ eV as a marker in the harmonic spectrum. We measure two-dimensional harmonic spectra in argon as a function of wavelength and angular divergence.
While we identify a clear CM in the spectrum when the target gas jet is placed after the laser focus, we find that the appearance of the CM varies with angular divergence and can even be completely washed out when the gas jet is placed closer to the focus. We also show that the argon CM appears at different wavelengths in harmonic and photo-absorption spectra measured under conditions independent of any wavelength calibration. We model the experiment with a simulation based on coupled solutions of the time-dependent Schr\"odinger equation and the Maxwell wave equation, thereby including both the single atom response and macroscopic effects of propagation and phase matching.   The single atom calculations confirm that the ground state of argon can be represented by its field free $p$ symmetry, despite the strong laser field used in high harmonic generation. Because of this, the CM structure in the harmonic spectrum can be described as the interference of continuum $s$ and $d$ channels, whose relative phase jumps by $\pi$ at the CM energy, resulting in a minimum shifted from the photoionization result. We also show that the full calculations reproduce the dependence of the CM on the macroscopic conditions. We calculate simple phase matching factors as a function of harmonic order and explain our experimental and theoretical observation in terms of the effect of phase matching on the shape of the harmonic spectrum.  Our results emphasize that phase matching must be taken into account to fully understand spectral features related to harmonic spectroscopy. Furthermore, we show that in some cases phase matching can be actively used to enhance the visibility of interference minima in high harmonic spectra.
\end{abstract}
\maketitle

\section{Introduction}
High harmonic generation (HHG) has the potential to image atomic and molecular electronic structures with Angstrom precision in the spatial domain and sub-femtosecond precision in the temporal domain. The principles behind the electronic imaging scheme are closely entangled with the mechanism of strong field harmonic generation itself \cite{Schafer_1993, Kulander_1993, Corkum_1993} which is described as a coherent splitting and recombination of an electronic wave function. The splitting of an initial bound electronic state $\psi_{\mbox{\tiny B}}$, of an atom or molecule occurs via tunnel ionization in a strong laser field. The resulting unbound electron wave function $\psi_{\mbox{\tiny UB}}$ is first accelerated away from the atomic core and subsequently accelerated towards the ion. As it reencounters the atomic core the time-dependent dipole matrix element, $d_z(t)= -e\langle\psi_{\mbox{\tiny UB}}|z|\psi_{\mbox{\tiny B}}\rangle$, gives rise to the emission of high harmonic radiation. Here we assume a laser field linearly polarized along the $\hat{z}$ direction.

Due to the coherent nature of the harmonic generation process, interferences between different channels in either the bound or the unbound electron wave function can occur in the dipole matrix element. These interference phenomena provide rich information about the electronic structure of the source medium. For example, the recombination radiation contains information about the internuclear distance of a diatomic source medium which can be described by a ``two center interference model" \cite{Lein_2002, Vozzi_2005}. Different ionization channels in the form of close lying molecular orbitals have also been found to lead to interference features in harmonic spectra \cite{McFarland_2008, Li_2008, Smirnova_2009, Haessler_2010}.

In this study we concentrate on a prototypical interference structure in atoms resulting from the recombination of different angular momentum channels in the unbound electron wave function. To characterize the angular momentum interference, we use the Cooper minimum \cite{Cooper_1962} in argon as a marker in the harmonic spectrum. Cooper minima are traditionally observed as a reduced photoionization cross section of many rare gas atoms and are clear signatures of the particular atomic ground state's nodal structure \cite{Samson_2002}. In argon the CM occurs at approximately 47 eV photon energy and results from a zero in the matrix element between the $d$ continuum waves and the argon $3p$ ground state, $\langle \psi_{d}(\epsilon_k) | x| \psi_{3p}\rangle$, at a particular value of the continuum energy $\epsilon_k$, a clear signature of the argon ground state's nodal structure. Neglecting the presence of the strong electric field during HHG, the recombination matrix element can be described as inverse photoionization. Thus the Cooper minimum should also occur in harmonic spectra. In this approximation, which can be checked against full calculations,  angular momentum selection rules allow for photoionization/recombination matrix elements with $d$ and $s$ channels in the unbound wave function.  In photoionization studies the $s$ and $d$ channels are incoherently superimposed due to the integration over the full solid angle \cite{Amusia_1990}. In contrast,  HHG is a coherent process and the laser field provides a strong directional sensitivity. Thus interference between $s$ and $d$ channel recombination radiation can alter the energetic position and modulation depth of the CM spectral marker from the $d$ channel in the full spectrum.

Amplitude modulations close to the position of the Cooper minimum have been observed in previous studies of argon high harmonic spectra using both  800~nm  \cite{Itatani_2004, Wahlstrom_1993, Zhou_1996, Shin_1999, AltucciC.:Depuma} and also longer mid-IR 2~$\mu$m fundamental wavelengths \cite{Colosimo_2008}. However, the observed amplitude minima have been addressed as a Cooper minumum only recently \cite{Minemoto_2008, Le_2008, Worner_2009}. Minemoto {\it et al.} simulate their HHG spectra consistently using the recombination matrix elements of a single field free argon atom. W\"orner {\it et al.} present the harmonic Cooper minimum as a test case for harmonic spectroscopy \cite{Worner_2009} by showing its independence with respect to laser intensity variations.

In this paper we extend the existing literature on interference phenomena in HHG by systematically studying the influence of phase matching on the CM in argon. Harmonic generation is necessarily a phase matched process, resulting in a well defined beam of harmonic radiation exiting the interaction region. Phase matching depends on many parameters and imposes a shape on the harmonic amplitude, which can mask the single atom response governed by the dipole recombination. We present experimental harmonic spectra showing the beam profile in addition to the spectral harmonic structure. We observe, in agreement with the existing literature, that the harmonic CM is shifted higher in energy with respect to the photoionization CM. Most importantly, we observe that the CM position and modulation depth is changed by phase matching conditions. In the most extreme case, the CM vanishes completely.

We simulate the experimental spectra using the solution of the time dependent Schroedinger equation (TDSE) for the single atom spectrum coupled with a Maxwell wave equation (MWE) propagation to explicitly include phase matching. From the single atom TDSE simulations we infer that the harmonic spectrum can be accurately approximated by $s$ and $d$ recombination channels only. This indicates that the ground state of argon is well described by its field free $p$ symmetry. The single atom calculations get close to the experimentally observed modulations. Thus, the phase matching, included in the experiment, conserves the underlying interference of $s$ and $d$ channels and in particular their relative phase. The MWE propagation reproduces the experimentally observed amplitude modulations at the harmonic CM. We calculate effective phase matching factors that explain why the single atom CM structure is enhanced in certain phase matching conditions, and suppressed in others. This sheds new light on phase matching as an additional parameter that can actively influence the harmonic spectrum and thus needs to be included when harmonic spectra are interpreted in terms of the electronic structure of the source medium. We show this for a angular momentum channel interference and we expect that the phase matching effects are generally true for other interference phenomena mentioned above.

\section{Experimental and theoretical framework}
We focus the output of a commercial Ti:Sapphire laser (pulse duration 30~fs, pulse energy 250 $\mu$J, central wavelength 800~nm) with a f=400~mm lens into a continuous flow gas jet. The harmonics between 20 and 70~eV pass through an Al filter (thickness 100~nm) onto a spherical grating. The dispersed image is then captured by an extreme ultraviolet (EUV) detector and image intensifier consisting of a bare microchannel plate (MCP) followed by a phosphor screen, which is viewed by a CCD camera. The spherical grating focuses only in the dispersion direction in the incidence plane of the EUV light (tangential direction) but keeps the natural divergence of the harmonic beam in the orthogonal (sagittal) direction. Since the beam hits the grating under grazing incidence, the finite size of the grating substrate acts like a slit that filters only the center part of the beam in the tangential direction.  The wavelength transmission function of the apparatus is taken into account in the data we present. We have also measured the wavelength dependent transmission of the Al filter in our lab in a separate experiment which takes the real oxidation of the filter into account. The wavelength transmission of the grating is provided by the manufacturer (Hitachi) and we estimate the MCP efficiency from ref. \cite{Hemphill_1997}. A camera scan through the focused laser beam results in a Rayleigh length of about 1.5~mm and a full width at half maximum in intensity of about 30~$\mu$m. The camera scan does not record the absolute laser focus position for the HHG experiment. We therefore estimate the laser focus position in the experimental setup by monitoring the laser plasma channel in a rare gas atmosphere.

To calculate the single atom harmonic spectrum we numerically integrate the TDSE using the single active electron (SAE) approximation. We use an $\ell$-dependent pseudo potential to describe the interaction between the active electron and the ion core. The pseudo potential is derived from an all-electron Hartree-Slater calculation and the $d$ channel potential has been adjusted so that the CM in the $p$-$d$ dipole matrix element is in the correct position (about 47 eV). This is done by the addition of a short range potential near the origin. We refer to Mauritsson {\it et al.} for details \cite{Mauritsson_2005}. When modifying the pseudo potential to give the correct position of the CM we take care that the scattering phase of the $d$ channel continuum functions is changed as little as possible.  The importance of this will become clear when we discuss the detailed shape of the HHG spectrum. We start from the $3p$, $m=0$ ground state wave function as the active electron and propagate forward in time \cite{SchaferBook_2008} to obtain the time-dependent wave function.
At each time step we calculate the acceleration form of the dipole moment as:
\begin{equation}
a(t) = -  \langle \psi(t) | [H,[H,z]] |\psi(t) \rangle,
\end{equation}
where $H$ is the full (atom plus laser field) Hamiltonian. The numerical calculation of the acceleration converges much more rapidly than the dipole itself, due to the fact that the dipole is proportional to z while the acceleration is
proportional to $z/r^3$ \cite{Burnett_1992}. In the frequency domain we can obtain the dipole form from the acceleration via the relation
$\tilde d(\omega) = e\,\tilde a(\omega)/\omega^2$. We then use $\tilde{d}(\omega)$ as the source term in our MWE solver.

In order to study the interference minimum in the HHG spectrum due to the $p$-$d$ CM we need to resolve the full HHG spectrum into separate  $s$ and $d$ continuum channel contributions. This can be done by calculating an approximate dipole moment assuming that the transitions that are responsible for the high harmonics are those that begin or end in the field free ground state. It has been shown that this is an excellent approximation for harmonics above the ionization threshold \cite{Krause_1992a}. In the case under consideration this means that the total time-dependent dipole
can be decomposed as $\bar{d}_z(t) = d_{s}(t) + d_{d}(t)$ where
\begin{eqnarray}
d_s (t) & = &-e  \langle \psi_s(t) | \phi_g \rangle \langle \phi_g |z| \psi_s(t) \rangle + c.c \nonumber \\
 d_d (t) & = & -e \langle \psi_d(t) | \phi_g \rangle \langle \phi_g |z| \psi_d(t) \rangle + c.c.
\label{dipole_sep}
\end{eqnarray}
In these expressions $\phi_g$ is the initial $3p$ state and $\psi_s(t)$ and $\psi_d(t)$ are the $\ell=0$ and $\ell=2$ components of the full time-dependent wave function. The approximate dipole $\bar{d}_z(t)$ converges much more rapidly than the full dipole because of the explicit use of the spatially compact ground state wave function in its calculation. A similar approximate expression for the acceleration form of the dipole in terms of $s$ and $d$ channel contributions can also be obtained. The agreement between the spectra derived from the full acceleration, the approximate acceleration and the approximate dipole is excellent and allows us to discuss the full spectrum as the coherent sum of two contributions that we can separately calculate. This allows us to calculate the phase between the two contributions and examine its effect on the CM, as we discuss below.

To calculate the macroscopic spectra for direct comparison with the experimental results, we have developed a numerical non-linear medium (NNLM) in which we solve the coupled SAE-TDSE and MWE  on a massively parallel supercomputing platform. We solve a uni-directional propagation equation in a frame that moves at the speed of light. In the frequency domain it takes the following form \cite{Brabec_2000}:
\begin{equation}
\nabla^{2}_\perp \tilde E(\omega) +\frac{2i\omega}{c}\frac{\partial \tilde E(\omega)}{\partial z}
=  - \frac{\omega^2}{\epsilon_0 c^2} \tilde P(\omega)
+ \frac{e^2}{\epsilon_0 m_e c^2} \widetilde{FT}[n_e(t)E(t)].
\label{MWE}
\end{equation}
The electric field $\tilde E(\omega)$ includes all the frequencies in the light field, both the initial laser frequency and the frequencies produced via harmonics generation and other nonlinear processes, and is related to $E(t)$ via a Fourier transform. The polarization field is $\tilde P(\omega) = 2n_0 \tilde d(\omega)$, where $n_0$ is the initial argon density and $\tilde d(\omega)$ is the Fourier transform of the time-dependent dipole moment resulting from the SAE-TDSE calculation above. The factor of two is due to the two $\ell=1, m=0$ electrons. We note that $\tilde P(\omega)$ includes both the linear and non-linear response of the atom to the multi-frequency field. The other source term on the right hand side is due to the space- and time-dependent plasma refractive index through the variation of the electron density $n_e(t)$. Note that we have omitted the dependence of $\tilde E(\omega)$, $\tilde P(\omega)$, $E(t)$, $n_0$ and $n_e$ on the cylindrical coordinates $r,z$ for notational simplicity.

We solve Eq.~(\ref{MWE}) by space marching through the argon gas, using the input laser spatiotemporal profile as the initial electric field. This is usually approximated as a focused Gaussian beam in space and a $\cos^4$ intensity envelope in time. At each step in the propagation direction we calculate the source terms in the time-domain via numerical integration of the TDSE as described above, for all points in the transverse direction. The time-dependent electron density is calculated from the density of the time-dependent wave function outside of a small sphere around the origin (see ref. \cite{Gaarde_2008} for details). We then Fourier transform the source terms to the frequency domain and use them to propagate to the next plane in the propagation direction.

The NNLM allows the description of  the evolution of the laser field during propagation, diffraction, focusing, and ionization-induced effects. In addition it describes  the generation, build-up, and phase matching of higher frequencies. Because of the way the polarization field is coupled to the electric field, the NNLM also includes effects usually described as linear, namely dispersion and absorption, within the SAE approximation. For more details on time-dependent and IR-assisted XUV absorption we refer to \cite{Gaarde_2010}.

To provide a qualitative gauge of the effects of macroscopic propagation, we also calculate approximate phase matching (PM) factors via the phase mismatch for harmonic q: $\Delta {\vec k}_q = {\vec k}_q - {\vec k}_{pol}$ as outlined by Balcou and collaborators \cite{Balcou_1997}. Here ${\vec k}_q$ is the wave vector of the propagating harmonic field, and ${\vec k}_{pol}$ is the wave vector of the newly generated harmonic field which depends on the phase variation of the source term (the non-linear polarization field), ${\vec k}_{pol} = {\vec \nabla } \phi_{pol}(r,z)$. Ignoring small effects of dispersion on the {\it harmonic} field we use  $k_q = q\omega_1/c$. The phase of the source term is given by
$\phi_{pol}(r,z) = q\phi_1(r,z) - \alpha_q I(r,z)$,
where $I(r,z)$ and $\phi_1(r,z)$ are the intensity and phase distribution of the driving field and $r$ and $z$ span the overlap of the laser focus with the non-linear medium. $\alpha_q$ is the phase coefficient of the intensity dependent phase for either the short or long trajectory, calculated from the position of the harmonic relative to the cutoff energy \cite{Lewenstein_1994, Lewenstein_1995}.
$\phi_1(r,z)$ is dominated by the geometric phase due to focusing. For the calculations in this paper we use $I(r,z)$ and $\phi_1(r,z)$ obtained from the full MWE-TDSE calculation which means that they also include the effect of free electron-induced defocusing and dispersion. Once we have obtained ${\vec k}_{pol}(r,z)$ we calculate the phase mismatch everywhere in the medium as
$\Delta k_q(r,z) = q\omega_1/c - |{\vec k}_{pol}(r,z)|$, and the angular divergence of the harmonic beam as $\theta(r,z) = \tan^{-1} (k_{pol,r}/k_{pol,z})$. To calculate the true phase matching factor for the harmonic radiation found at divergence $\theta$ in the far field, one would then have to integrate all the phase mismatch contributions with divergence $\theta$ in the near field over the length of the medium, which is very complicated because there is not a one-to-one mapping of the near field $r$ to $\theta(r)$. Instead we approximate the PM factor for radiation with angular divergence $\theta$ as:
\begin{equation}
F(\theta(r), z_L/2) = {\rm sinc}^2 [\Delta k_q(r,z_L/2)z_L/4]\ I^4(r,z_L/2).
\end{equation}
That is, we evaluate the PM factor at the specific plane $z=z_L/2$, where $\pm z_L/2$ represents the half-width at half-maximum of the density profile of the argon gas. The full calculation shows that the harmonic radiation predominantly builds up in the latter half of the medium. The PM factor has been weighted with $I^4(r)$ in order to approximate the intensity dependence of the harmonic.

\section{Results and Discussion}

\begin{figure}\centering
\includegraphics[width=15 cm]{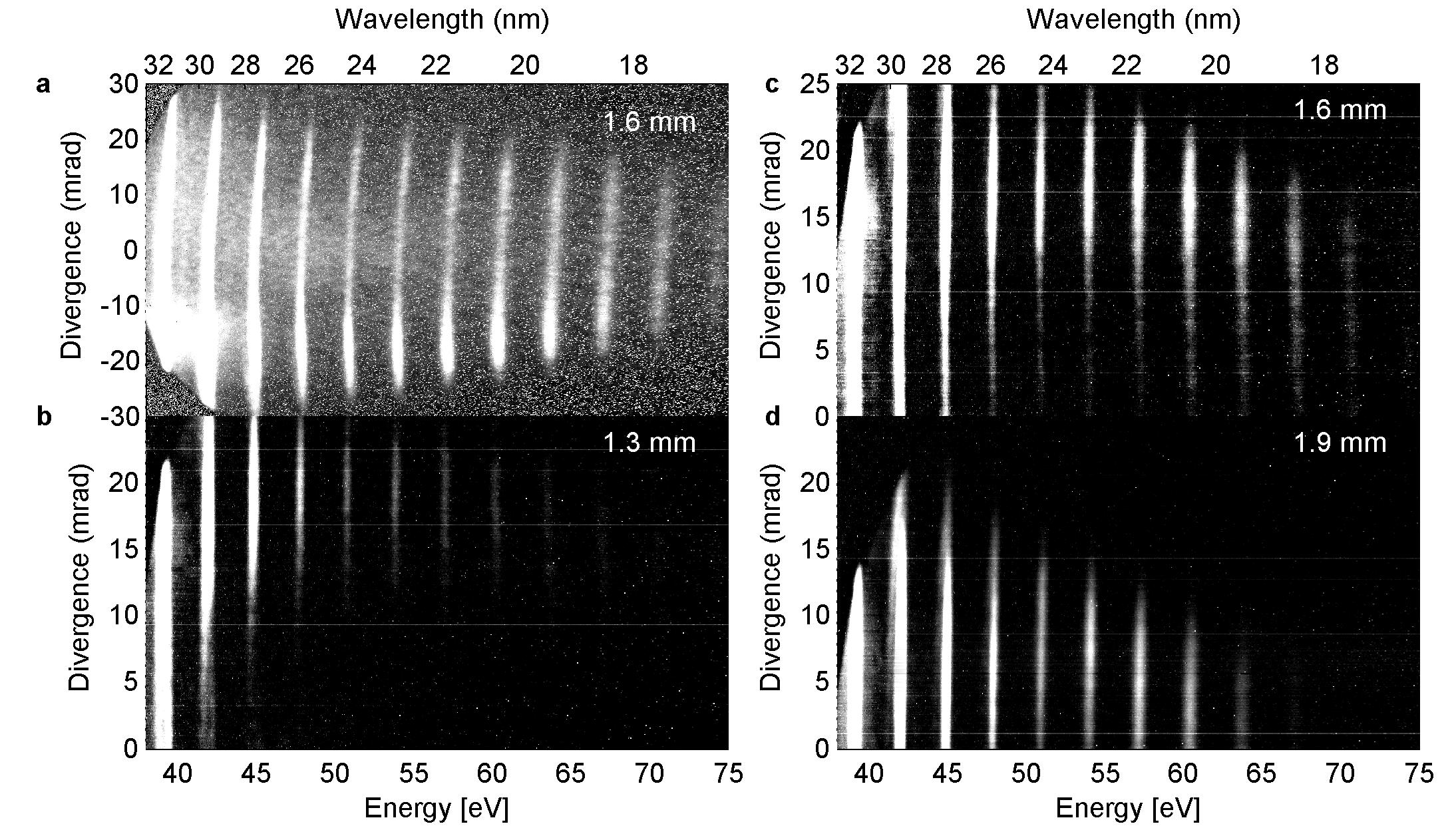}
 \caption {Experimental harmonic spectra of Ar dispersed by the spherical grating for different gas jet positions with respect to the laser focus. The y axes represent divergence and the x axes represent wavelength. a) Spectrum not corrected for any apparatus response. One clearly identifies the individual harmonics that are modulated along the divergence direction. b-d) Spectra corrected for spectral MCP response, grating reflectivity and Al-filter transmission. The spectra are symmetrized in the divergence coordinate. As the gas jet is scanned through the focus, the shape of the harmonic spectrum is described first by strong off-axis contributions (b), turning into a ''hole" on axis (c) which is filled as the divergence shrinks (d). The Cooper minimum is clearly visible from 22 to 24~nm on axis in (c) and (d) while in (d) the off axis contribution also shows a Cooper minimum.}\label{Cooper2d}
\end{figure}

Figure~\ref{Cooper2d}(a) shows a 2D far field HHG spectrum from argon, measured with the gas jet placed 1.6 mm after  the laser focus. This spectrum is {\it not} corrected for any apparatus response. The harmonics are modulated in the divergence direction orthogonal to the spectral axis. The general asymmetry between positive and negative divergence most likely results from a non perfect laser beam profile and slight misalignment in the spectrometer. We can clearly identify a minimum in the harmonics on axis along the zero of the divergence coordinate. Towards higher divergence, this minimum vanishes. To quantify our spectral envelopes, we have corrected the data for the MCP, grating and Al-filter responses. We have also symmetrized the spectra with respect to the divergence by averaging the positive and negative divergence contributions. Figures~\ref{Cooper2d}(b)-(d) show 2D grayscale plots of the experimental harmonic spectra when the gas jet is placed 1.3~mm, 1.6~and at 1.9~mm after the laser focus, respectively. The spectrum in Fig.\ref{Cooper2d}(c) is the processed version of the raw data in Fig.\ref{Cooper2d}(a) and displays the same angular variation of the spectral shape: a broad spectral minimum is visible on-axis centered around 53 eV, whereas the off-axis radiation does not exhibit an obvious minimum. As the gas jet is moved closer to the focus [Fig.~\ref{Cooper2d}(b)] the spectrum becomes dominated by the off-axis contribution which also does not display a clear minimum. Further away from the focus [Fig.~\ref{Cooper2d}(d)] we observe a spectral minimum both on-axis and off-axis. It is interesting to note that in this jet position, the minimum is centered on 51 eV.

\begin{figure}\centering
\includegraphics[width=10 cm]{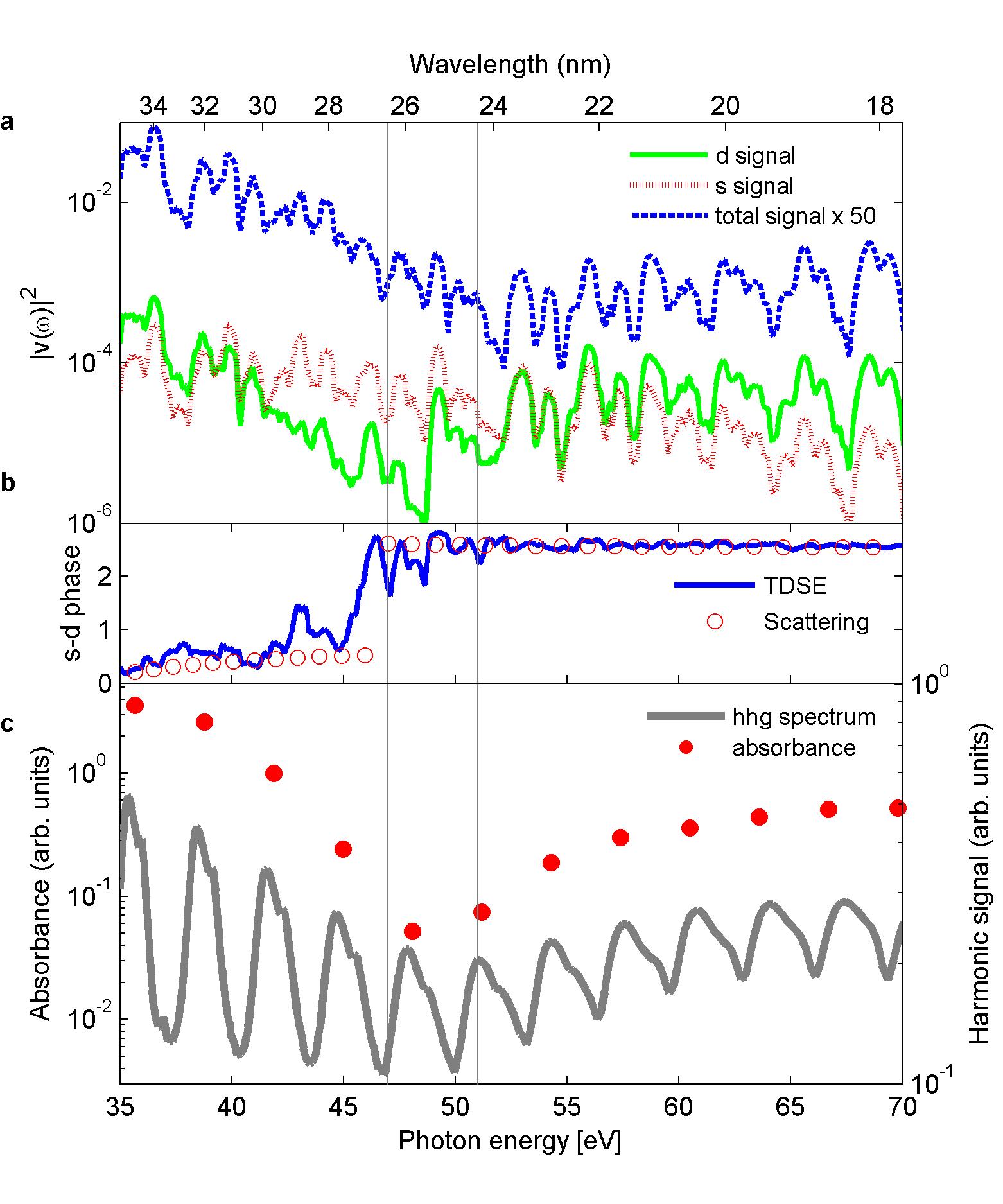}
 \caption {a) Single atom harmonic spectra calculated using the approximate dipole moment. We show the total (solid), and the $s$-  and $d$-contributions (dashed and dotted) separately. b) Phase between the calculated s and d channel dipoles jumping from 0 to $\pi$ near the Cooper minimum of the d channel, calculated via the TDSE (line) and the scattering phase approach (symbols). c) Absorption spectrum of argon (solid circles) and harmonic spectrum taken in the same calibrated spectrometer. The minimum in the harmonic spectrum is blueshifted compared to the absorption spectrum by one harmonic.}
 \label{Exp_single_atom}
\end{figure}

To explore the position and appearance of the  CM in the HHG spectrum,  we show in Fig.\ref{Exp_single_atom}(a) calculated single atom harmonic spectra for argon generated by a 780~nm, 35~fs pulse with a peak intensity of $3.5\times 10^{14}$~W/cm$^2$. We plot $|v(\omega)|^2 = |\omega d(\omega)|^2$
in analogy with the macroscopic response, since Eq.~(\ref{MWE}) for a plane wave and perfect phase matching would yield an electric field which is proportional to $\omega/c$ times the polarization field. The spectra have been smoothed  by using a moving average of 0.4~eV.
A clear minimum is visible in the spectrum around 52~eV. We also show the separately calculated contributions to the harmonic spectrum from $s$- and $d$-recombinations to the ground ($p$) state.We note that, as often happens when constructing single electron pseudo potentials, the ground state to $s$ continuum photoionization matrix element, $\langle \phi_g |z |\psi_{s}(\epsilon_k)\rangle$, is larger than the same matrix element calculated using the all electron potential from which the pseudo potential was derived. This deviation is fairly constant over the spectral region of interest in our calculations. We take the opportunity of
having separate $s$ and $d$ contributions in the approximate dipole to correct for this by scaling the $d_s(t)$
contribution by a constant factor of 0.3. Looking at the separate $s$ and $d$ contributions we see that the  $s$-contribution is mainly unstructured, whereas the $d$-contribution shows a deep minimum around 47~eV, in agreement with photo-absorption measurements (see for instance \cite{Samson_2002} and Fig.~\ref{Exp_single_atom}(c)).
This shows that the position of the CM in the $p\!-\!d$ recombnination channel is unaffected by the strong field.

The total spectrum in Fig.~\ref{Exp_single_atom}(a) results from the coherent addition of the $s$ and $d$ contributions. Thus both the amplitudes and phases of the two contributions determine the shape of the full spectrum. In Fig.~\ref{Exp_single_atom}(b) we show the relative phase between the $s$- and $d$-contributions, calculated by:
\begin{equation}
cos(\phi_{s-d}(\omega)) = \frac{|d_{total}(\omega)|^2 - |d_{s}(\omega)|^2 - |d_{d}(\omega)|^2 }{2|d_{s}(\omega)||d_{d}(\omega)|}, \label{relPhase}
\end{equation}
where $|d(\omega)|^2$ are the single atom spectra shown in Fig.~\ref{Exp_single_atom}(a). The relative phase exhibits a jump of $\approx \pi$ at the position of the CM, consistent with a sign-change of the field-free $d\!-\!p$ recombination matrix element. The numerical result is very sensitive to small oscillations of the dipole strength, which can slightly shift the appearance of the phase jump. Because the $s$ and $d$ contributions were relatively in phase for energies below the CM,  they will be almost out of phase for energies above it. This means that the minimum in the coherently added total harmonic spectrum will occur not at the position of the $d$-channel CM (where the $s$-contribution is non-zero), but at the energy where the $s$ and $d$-contributions have equal magnitude.  Thus the Cooper minimum is shifted towards higher harmonics where the $s$ and $d$ amplitudes are approximately equal. The exact position of this minimum, which is shallower than than $d$-channel CM, is sensitive to the relative shape of the $s$- and $d$-contributions and can move a few eV depending on the intensity.

In recent work \cite{Le_2008}, C.D. Lin and coworkers have presented a ``quantitative  rescattering theory" (QRS) of HHG which posits that the phase of the recombining electron wave function can be separated into two energy-dependent contributions (see also ref.~\cite{Lin_2010} for a review of QRS and molecular HHG). The first contribution is an intensity-dependent phase due to the propagation of the electron wave packet created by tunnel ionization. This phase applies to all angular momentum channels equally, and is responsible for the ``long'' and ``short'' trajectory behavior familiar from many studies of HHG. The second phase contribution is due to the recombination matrix element and is taken to be equal to the energy-dependent scattering phase of the field-free continuum wave function with kinetic energy $\epsilon_k$ such that $k^2/2 +I_p =  \omega_q$ where $I_p$ is the ionization potential and $\omega_q$ is the energy of the $q^{\rm th}$ harmonic. This phase is channel-dependent
but intensity-independent. We find, using our approximate dipole calculation, that the inter-channel phase is indeed intensity-independent. Moreover, we can compare the inter-channel phase to the phase predicted by QRS see equation 33 of reference \cite{Le_2009a} by calculating the scattering phase shifts for the continuum wave functions in our pseudo potential and adding the CM phase jump by hand at the appropriate energy. We get a relative phase in agreement with that in Fig.~\ref{Exp_single_atom}(b). This result justifies the use of scattering phase shifts for the simulation of HHG spectra around the argon CM as done in refs. \cite{Minemoto_2008, Worner_2009}.
We remark that the presence of a deep interference minimum is strongly dependent on the phase behavior shown in Fig.~\ref{Exp_single_atom}(b). If the relative contributions to the total dipole were not relatively in or out of phase near the CM then the $\pi$ phase jump would not yield the behavior shown in Fig.~\ref{Exp_single_atom}(a). It remains to be seen if CM in other rare gas atoms will yield the same behavior.

Fig.~\ref{Exp_single_atom}(c) presents an experimental demonstration of the relative shift of the CM between the $d$-recombination matrix element and the harmonic spectrum. We show an  absorption spectrum of argon overlapped with a harmonic spectrum. The absorption spectrum was recorded by using harmonics generated in N$_2$ as the spectral source, and a EUV spectrometer chamber  flooded with argon as an absorption cell. For this purpose we used a different spectrometer, equipped with a CCD detector, than the one used in the measurements shown in Fig.~\ref{Cooper2d}. This spectrometer has been carefully calibrated by plasma emission lines as described in  \cite{Farrell_2009} and the harmonic spectrum in Fig.~\ref{Exp_single_atom}(c) was recorded in the same calibrated spectrometer. We see a characteristic broad minimum around 47~eV in the photoabsorption spectrum. It agrees perfectly with photoionization spectra recorded at synchrotron sources \cite{Samson_2002}, and with the minimum in the theoretical $d$-contribution spectrum in Fig.~\ref{Exp_single_atom}(a). Also in agreement with the theory is the shift to higher energy of the harmonic CM.

Phase matching and other macroscopic effects also play a role in shaping the total spectrum and in the position and appearance of the harmonic CM. To explain the full set of 2d spectra in Fig.~\ref{Cooper2d} and their change of the CM with jet-position, we now turn to full simulations of HHG including phase matching.

\begin{figure}\centering
\includegraphics[width=15 cm]{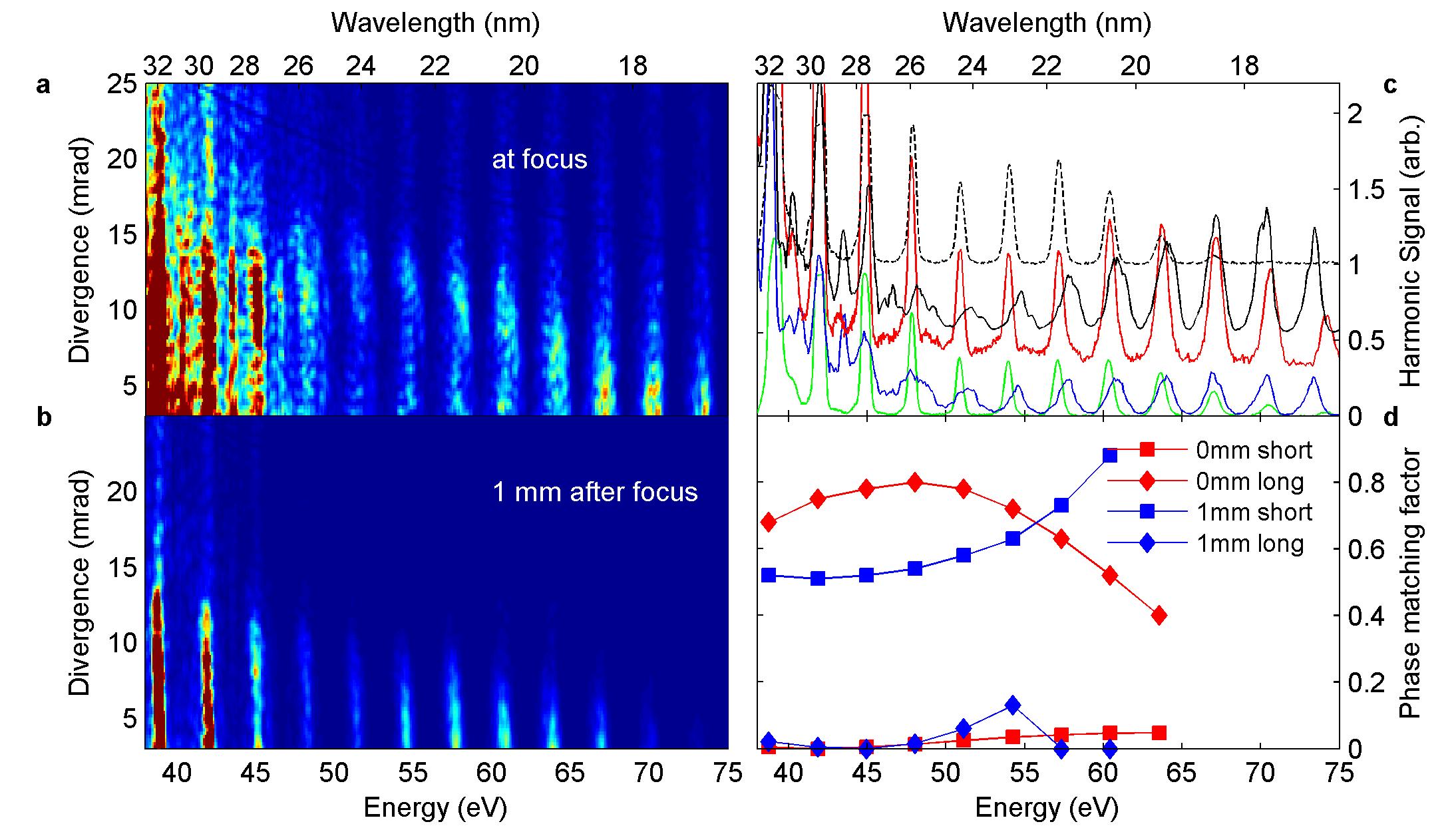}
 \caption {Calculated harmonic spectra as a function of divergence and harmonic number for different gas jet positions at and after the laser focus in the direction towards the spectrometer. a) at the focus, b) 1~mm behind the focus. c) Comparison of integration at the center and the outside of the experimental spectrum in Fig.~\ref{Cooper2d}(c) with the respective integrations of the theoretical spectrum in Fig.~\ref{CooperPhasematch}(b). From bottom to top: large divergence experiment (green) and theory (blue), small divergence experiment (red) and theory (black). The black dashed line is the spectrum for small divergence from Fig.~\ref{Cooper2d}(d). d) Phase matching factors (see text).
 }\label{CooperPhasematch}
\end{figure}

Fig.~\ref{CooperPhasematch} shows 2D harmonic spectra resulting from the full TDSE-MWE calculations, for different gas jet positions in analogy to the experimental spectra shown in Fig.~\ref{Cooper2d}. We observe the same trend as in the experimental spectra: When the gas jet is close to the focus [Fig.~\ref{CooperPhasematch}(a)]  the small divergence part of the harmonic spectrum exhibits a Cooper minimum analogous to that in Fig.~\ref{Cooper2d}(c). The large divergence region of the spectrum exhibits a general decay of the signal towards higher harmonics without a pronounced minimum. Fig.~\ref{CooperPhasematch}(c) shows a comparison of experimental and theoretical spectra integrated over the low divergence and high divergence regions in Fig.~\ref{Cooper2d}(c) and Fig.~\ref{CooperPhasematch}(a). The experiment and theory show excellent agreement. In order to directly compare the experimental and theoretical signals, we have multiplied the theoretical traces by the square of the photon energy. This accounts for the fact that the experimental spectrum is measured in the wavelength domain whereas the simulations are performed in the energy domain. When the gas jet is moved downstream from the focus [Fig.~\ref{CooperPhasematch}(b)] the harmonic CM is clearly visible both on-axis and off-axis, in very good agreement with the experimental spectrum in Fig.~\ref{Cooper2d}(d). In addition, both theory and experiment show the harmonic CM to be broader and located at slightly higher energy (by a few eV) when the gas jet is closer to the focus.

The experiment and the calculations differ somewhat on the absolute positions of the gas jet and the laser focus in which we observe the two different behaviors discussed above. In particular, the effective focal length in the experiment seems larger than that in the theory. We attribute this to the imperfect modeling of the laser beam, which in the calculation is assumed to be a Gaussian transverse electromagnetic mode (TEM) 00. Also, the density used in the experiment is not known exactly which means that ionization induced defocusing could play a larger or smaller role than in the calculation.

Besides the comparison of the integrations from theory and experiment, we compare the spectra integrated over small divergence from Fig.~\ref{Cooper2d}(c)(red line) and Fig.~\ref{Cooper2d}(d)(blue line). Both spectra clearly show a harmonic CM above the CM energy from photoionization experiments. However, we observe a strong difference in the shape and the exact appearance of the minimum. The measurement at 1.6~mm after the focus (red line) has a wide minimum stretching over three harmonics with almost equal amplitude. The middle of the minimum is located at 54~eV and also this particular harmonic is slightly lower than the surrounding ones. In contrast, the measurement at 1.9~mm after the focus (dashed line) we identify a very sharp CM at 51~eV. Although the CM has been observed to be independent of the laser intensity at fixed jet position \cite{Worner_2009}, the CM does seem to be influenced by moving the jet. Two different effects could contribute to this observation. First, the phase matching is influencing the shape of the CM and due to its relatively shallow modulation depth the exact CM position is easily shifted. We will expand on this argument further below. Second, our single atom calculations shown in Fig.~\ref{Exp_single_atom}(a) show a blueshift of the CM position with increasing intensity in the total signal. The CM position in the d channel is not dependent on intensity and thus the shift in the total channel is due to the different relative strength of the d and s channel. The results of Minemoto and collaborators \cite{Minemoto_2008} support our observation of the sensitivity of the CM to the generating conditions. They observed the CM in argon at two different energies (approx. 48 eV and 57 eV) when generating harmonics with two different laser sources.

The strong agreement between experiment and theory can be understood qualitatively by considering the underlying mechanism of HHG spectral shaping by phase matching. The hole in the argon spatio-spectral distribution is due to different phase matching conditions for the short and long trajectory contributions to the harmonic dipole moment \cite{Salieres_1995,  Bellini_1998}. In particular, in our high-intensity, tight-focusing geometry the contribution from the long trajectory becomes annular for a range of harmonics in the plateau when the focus is placed in the center of the jet. Furthermore, phase matching strongly prefers the long trajectory contribution over the short trajectory contribution for these harmonics. This is illustrated in Fig.~\ref{CooperPhasematch}(d) in which we show approximate PM factors calculated (as described in the previous section)  as a function of harmonic order for the conditions used in the full calculation. The long trajectory phase matching is mostly influencing the off axis spectrum, whereas the short trajectories are mostly determining the on axis spectral intensities.

The red symbols in Fig.~\ref{CooperPhasematch}d show the phase matching conditions close to the focus. At this position the long trajectory phase matching dominates. The long trajectory PM factor first increases, is maximized close to the position of the harmonic CM, and then decreases toward the cutoff energy. The short-trajectory PM factor (which is only relevant for small divergences) increases over the entire spectral range. This means that the on-axis appearance of the CM is enhanced since the phase matching compensates for the trend of the spectrum to decrease towards the cutoff energy. The off-axis appearance of the CM is suppressed by the PM factor which leads to an enhancement of the harmonic yield around the CM energy. Therefore one needs to apply care when interpreting spectra that are integrated over the whole divergence range (as generally recorded with toroidal grating spectrometers). The strong off-axis contributions can easily wash out the clear CM on axis.
The blue symbols in Fig.~\ref{CooperPhasematch}d show the phase matching conditions after the laser focus. They illustrate that phase matching can have an important effect on the appearance of the CM: both the short and long trajectory PM factors increase sharply as the harmonic order increases. This enhances the recovery of the spectral strength after the CM and leads to the very clear appearance of the CM in this focusing configuration.

Finally, we note that although phase matching changes the overall shape of the harmonic spectrum, it does little to change the relative shape of the $s$- and the $d$-contributions to the spectrum (such as shown in Fig.~\ref{Exp_single_atom}(a)). In other words, both the $s$-contribution and the $d$-contribution are changed in similar ways by phase matching. We can calculate the ``macroscopic'' relative phase between the two contributions by inserting into Eq.~(\ref{relPhase}) the macroscopic, radially integrated spectra emerging from the gas jet. This phase is remarkably similar to the single atom relative phase shown in Fig.~\ref{Exp_single_atom}(b) and does not change with intensity or focusing conditions.

\section{Summary}
We have shown experimental harmonic spectra of argon that are resolved with respect to both wavelength and divergence. These 2d spectra are made possible by use of a spherical grating, only focusing the harmonics in the tangential dimension. We compare the 2d spectra to Ar photoabsorption spectra and find a harmonic Cooper minimum which is located several eV above the photoabsorption Cooper minimum. The detailed shift and modulation depth of the harmonic Cooper minimum depend critically on gas jet position and divergence of the harmonics. Close to the laser focus, the off-axis harmonics dominate and no clear Cooper minimum can be distinguished. A little further away in the direction of the laser propagation direction, we observe a broad Cooper minimum for small divergence (on axis), whereas the off-axis contributions show a slight knee. About 2~mm after the focus, we identify a Cooper minimum for all divergences.

We simulate our data using coupled solutions of the time dependent Schr\"odinger equation for the single atom response and the Maxwell wave equations to include phase matching. We decompose the harmonic spectrum into contributions from the $s$ and $d$ continua, which reproduces the full TDSE spectrum with high accuracy. From that we deduce that the ground state of Ar maintains its field free $p$ shape despite the presence of the strong laser field in HHG. Because of this, the harmonic spectrum can be described as an interference of continuum $s$ and $d$ channels, whose relative phase jumps from 0 to $\pi$ at the $d$ channel Cooper minimum. This predicts the shift of the harmonic Cooper minimum with respect to the photoabsorption measurements. We also find that the Cooper minimum position in the $d$ channel, as well as the relative phase between $s$ and $d$ channel, is insensitive to the chosen field intensity. The CM position in the simulated harmonic spectrum can vary slightly depending on the intensity. The full phase matching simulations reproduce the divergence features of the experiments and their dependence on the relative position of gas jet and laser focus.

In conclusion, we have observed that different phase matching conditions can alter the shape of the single atom spectrum. In one case they wash it out, in a more fortunate case they enhance weak features. Generally, high harmonic spectroscopy is performed for gas jet positions after the focus similar or equal to our spectra shown in Fig.~\ref{Cooper2d}d. It is generally stated that this position is ``neutral" with respect to macroscopic effects, meaning that the phase matching factors are flat. Our simulations and their agreement with the experimental spectra however indicate, that even for these particular phase matching conditions, the single atom response is still modulated by phase matching; fortunately in this case in the form of a general enhancement of the structural minimum.

\acknowledgements
We thank the Department of Energy, Office of Basic Energy Science, AMOS program for support through the Stanford PULSE Institute. L.~S.~S. acknowledges the support of a NDSEG fellowship.
This work was supported by the National Science Foundation under grant~Nos.~PHY-0449235 (MG) and PHY-0701372 (KS). We also gratefully acknowledge support from the PULSE Institute at Stanford University (MG and KS). High performance computational resources were provided by the Louisiana Optical
Network Initiative, www.loni.org.

\end{document}